\documentclass[11pt,twoside]{article}
\usepackage{asp2014}
\aspSuppressVolSlug
\resetcounters
\begin{document}
\title{The Online Observation Quality System Implementation for the ASTRI Mini-Array Project}
\author{L. Castaldini$^1$, N. Parmiggiani$^1$, A. Bulgarelli$^1$, L. Baroncelli$^1$, V. Fioretti$^1$, A. Di Piano$^1$, I. Abu$^1$, M. Capalbi$^6$, O. Catalano$^6$, V. Conforti$^1$, M. Fiori$^5$, F. Gianotti$^1$, F. Lucarelli$^{2,3}$, M. C. Maccarone$^6$, T. Mineo$^6$, S. Lombardi$^2$, V. Pastore$^1$, F. Russo$^1$, P. Sangiorgi$^6$, S. Scuderi$^7$, G. Tosti$^{4,8}$, M. Trifoglio$^1$, and L. Zampieri$^5$ for the ASTRI Project$^9$.}
\affil{$^1$INAF/OAS Bologna, Via P. Gobetti 93/3, I-40129 Bologna, Italy.}
\affil{$^2$INAF/OAR Roma, Via di Frascati 33, I-00078 Monte Porzio Catone, Roma, Italy.}
\affil{$^3$ASI/SSDC Roma, Via del Politecnico snc, I-00133 Roma, Italy.}
\affil{$^4$Universit\`{a}' degli Studi di Perugia, Dip.to di Fisica e Geologia, Via A. Pascoli, I-06123 Perugia, Italy.}
\affil{$^5$INAF/OAPd, Vicolo Osservatorio 5, I-35122 Padova, Italy.}
\affil{$^6$INAF/IASF Palermo, Via Ugo La Malfa 153, I-90146 Palermo, Italy.}
\affil{$^7$INAF/IASF Milano, Via Alfonso Corti 12, I-20133 Milano, Italy.}
\affil{$^8$INAF/OA Brera, via Brera 28, I-20121 Milano, Italy.}
\affil{$^9$http://www.astri.inaf.it/en/library}
\paperauthor{Luca~Castaldini}{luca.castaldini@inaf.it}{0009-0000-5501-4328}{INAF/OAS~Bologna}{}{Bologna}{Bologna}{40055}{Italy}
\paperauthor{Nicol\`{o}~Parmiggiani}{nicolo.parmiggiani@inaf.it}{0000-0002-4535-5329}{INAF/OAS~Bologna}{}{Bologna}{Bologna}{40129}{Italy}
\paperauthor{Andrea~Bulgarelli}{andrea.bulgarelli@inaf.it}{0000-0001-6347-0649}{INAF/OAS~Bologna}{}{Bologna}{Bologna}{40129}{Italy}
\paperauthor{Leonardo~Baroncelli}{leonardo.baroncelli@inaf.it}{0000-0002-9215-4992}{INAF/OAS~Bologna}{}{Bologna}{Bologna}{40129}{Italy}
\paperauthor{Valentina~Fioretti}{valentina.fioretti@inaf.it}{0000-0002-6082-5384}{INAF/OAS~Bologna}{}{Bologna}{Bologna}{40129}{Italy}
\paperauthor{Ambra~Di Piano}{ambra.dipiano@inaf.it}{0000-0002-9894-7491}{INAF/OAS~Bologna}{}{Bologna}{Bologna}{40129}{Italy}
\paperauthor{Ismam~Abu}{ismam.abu@inaf.it}{0009-0001-7973-8192}{INAF/OAS~Bologna}{}{Bologna}{Bologna}{40129}{Italy}
\paperauthor{Milvia~Capalbi}{milvia.capalbi@inaf.it}{0000-0002-9558-2394}{INAF/IASF~Palermo}{}{Palermo}{Palermo}{90146}{Italy}
\paperauthor{Osvaldo~Catalano}{osvaldo.catalano@inaf.it}{0000-0002-9554-4128}{INAF/IASF~Palermo}{}{Palermo}{Palermo}{90146}{Italy}
\paperauthor{Vito~Conforti}{vito.conforti@inaf.it}{0000-0002-0007-3520}{INAF/OAS~Bologna}{}{Bologna}{Bologna}{40129}{Italy}
\paperauthor{Michele~Fiori}{michele.fiori@inaf.it}{0000-0002-7352-6818}{INAF/OAPd}{}{Padova}{Padova}{35122}{Italy}
\paperauthor{Fulvio~Gianotti}{fulvio.gianotti@inaf.it}{0000-0003-4666-119X}{INAF/OAS~Bologna}{}{Bologna}{Bologna}{40129}{Italy}
\paperauthor{Fabrizio~Lucarelli}{fabrizio.lucarelli@inaf.it}{0000-0002-6311-764X}{INAF/OAR~Roma}{}{Monte~Porzio~Catone}{Roma}{00078}{Italy}
\paperauthor{Maria Concetta~Maccarone}{cettina.maccarone@inaf.it}{0000-0001-8722-0361}{INAF/IASF~Palermo}{}{Palermo}{Palermo}{90146}{Italy}
\paperauthor{Teresa~Mineo}{teresa.mineo@inaf.it}{0000-0002-4931-8445}{INAF/IASF~Palermo}{}{Palermo}{Palermo}{90146}{Italy}
\paperauthor{Saverio~Lombardi}{saverio.lombardi@inaf.it}{0000-0002-6336-865X}{INAF/OAR~Roma}{}{Monte~Porzio~Catone}{Roma}{00078}{Italy}
\paperauthor{Valerio~Pastore}{valerio.pastore@inaf.it}{0000-0002-4776-5890}{INAF/OAS~Bologna}{}{Bologna}{Bologna}{40129}{Italy}
\paperauthor{Federico~Russo}{federico.russo@inaf.it}{0000-0002-3476-0839}{INAF/OAS~Bologna}{}{Bologna}{Bologna}{40129}{Italy}
\paperauthor{Pierluca~Sangiorgi}{pierluca.sangiorgi@inaf.it}{https://orcid.org/0000-0001-8138-9289}{INAF/IASF~Palermo}{}{Palermo}{Palermo}{90146}{Italy}
\paperauthor{Salvatore~Scuderi}{salvatore.scuderi@inaf.it}{0000-0002-8637-2109}{INAF/IASF~Milano}{}{Milano}{Milano}{20133}{Italy}
\paperauthor{Gino~Tosti}{gino.tosti@unipg.it}{0000-0002-0839-4126}{Universit\`{a}~degli~Studi~di~Perugia}{}{Perugia}{Perugia}{06123}{Italy}
\paperauthor{Massimo~Trifoglio}{massimo.trifoglio@inaf.it}{0000-0002-2505-3630}{INAF/OAS~Bologna}{}{Bologna}{Bologna}{40129}{Italy}
\paperauthor{Luca~Zampieri}{luca.zampieri@oapd.inaf.it}{0000-0002-9286-7693}{INAF/OAPd}{}{Padova}{Padova}{35122}{Italy}
\begin{abstract}
The ASTRI Mini-Array project, led by the Italian National Institute for Astrophysics, aims to construct and operate nine Imaging Atmospheric Cherenkov Telescopes for high-energy gamma-ray source study and stellar intensity interferometry. Located at the Teide Astronomical Observatory in Tenerife, the project's software is essential for remote operation, emphasizing the need for prompt feedback on observations. This contribution introduces the Online Observation Quality System (OOQS) as part of the Supervisory Control And Data Acquisition (SCADA) software. OOQS performs real-time data quality checks on data from Cherenkov cameras and Intensity Interferometry instruments. It provides feedback to SCADA and operators, highlighting abnormal conditions and ensuring quick corrective actions for optimal observations. Results are archived for operator visualization and further analysis. The OOQS data quality pipeline prototype utilizes a distributed application with three main components to handle the maximum array data rate of 1.15 Gb/s. The first is a Kafka consumer that manages the data stream from the Array Data Acquisition System through Apache Kafka, handling the data serialization and deserialization involved in the transmission. The data stream is divided into batches of data written in files. The second component monitors new files and conducts analyses using the Slurm workload scheduler, leveraging its parallel processing capabilities and scalability. Finally, the process results are collected by the last component and stored in the Quality Archive.
\end{abstract}

\section{Introduction}
The \textbf{ASTRI Mini-Array} \citep{SCUDERI202252} project deals with the development of the next-generation Imaging Air Cherenkov Telescopes for ground-based gamma-ray astronomy. The ASTRI Mini Array observatory is located at Mount Teide (Tenerife, Spain), and the telescopes are currently under construction. ASTRI Mini-Array comprises nine telescopes performing high-sensitivity gamma-ray observations in the energy range between 1 and several hundred TeV. These observations rely on "air showers" detection, the result of energetic photons interacting with Earth's atmosphere, generating UV-blue light, known as Cherenkov radiation. The Cherenkov cameras produce different telemetry packets at high frequency and bandwidth. Scientific packets are generated as the camera detects events, resulting in a continuous stream of 16 kB-sized packets peaking in frequency at 1 kHz, leading the whole array data rate to 1.15 Gb/s. To ensure the highest possible data quality, ASTRI features the \textit{Online Observation Quality System} (OOQS), the software system performing data quality checks on camera data in real-time.
\section{Scada Context and OOQS Architecture}
The OOQS is part of the \textit{Supervisory Control And Data Acquisition} (SCADA) \citep{10.1117/12.2629164}, the software deployed at the Mini-Array Observatory controlling the operations carried out onsite. SCADA comprises other systems like \textit{Central Control System} (CCS),\textit{ Telescope Control System} (TCS), \textit{Array Data Acquisition System} (ADAS) and \textit{Monitoring and Alarm system}, as shown in Fig.\ref{OOQS_SCADA_diag}.
\articlefigure{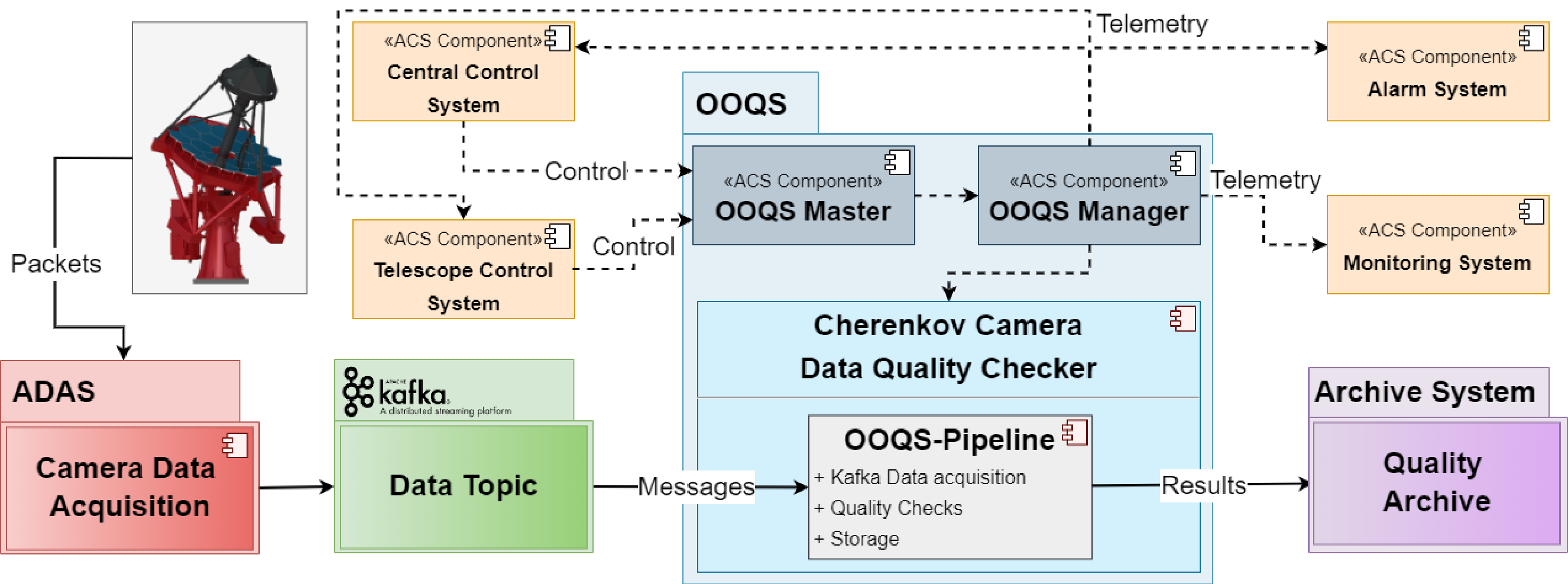}{OOQS_SCADA_diag}{The context of OOQS in SCADA. ADAS acquires packets and streams messages to the OOQS with Kafka; data quality results are stored in the Quality Archive. ACS components control the OOQS lifecycle, handle alarms, and share the observation quality status.}
OOQS receives camera data and conducts fast analyses of the \textit{data quality} (dq) to allow SCADA to take automated corrective actions and notify the nightly operator supervising the observation if the data acquired by the telescopes does not satisfy the quality requirements. 
OOQS interfaces with ADAS \citep{10.1117/12.2626600}, the system responsible for acquiring camera data, using Apache Kafka\footnote{Apache Kafka is a distributed event streaming platform \url{https://kafka.apache.org/}} to receive the camera packets. The data transmission involves encoding with Apache Avro\footnote{Apache Avro is a serialization format suitable for complex data 
types \url{https://avro.apache.org/}} \citep{Parmiggiani_2022}. The \textit{Cherenkov Camera DQ Checker}, which is a part of OOQS software, contains two quality checkers: Scientific and Variance. 
The Scientific Checker samples camera images and computes histograms on camera trigger number and pixel ADC values, checking if they are within a defined range. The Variance Checker compares the ratio between high-gain and low-gain of the ADC values and checks the telescope pointing deviation for corrective action. The dq results are stored permanently in the Quality Archive.
OOQS interfaces with other SCADA sub-systems through ALMA Common Software\footnote{A software infrastructure based on a distributed Component-Container model, where components are implemented as CORBA objects in various programming languages \url{https://confluence.alma.cl/display/ICTACS}} (ACS). OOQS connects with SCADA with two ACS components: Master and Manager.
CC starts the OOQS Master, while TCS configures and controls its operativity during the data capture. The OOQS Manager controls the DQ Checkers and provides feedback on the observation status to the TCS, CCS, Monitoring, and Alarm system.
\section{OOQS Pipeline}
The \textbf{OOQS pipeline}, shown in Fig.\ref{OOQS_Pipeline_diagram}, is the software performing the data processing required by the Cherenkov Camera DQ Checker. 
It receives data from a Kafka Topic, conducts quality checks, consolidates results, and stores them in a database. It consists of three synchronized daemons triggered by file system events. 
\articlefigure{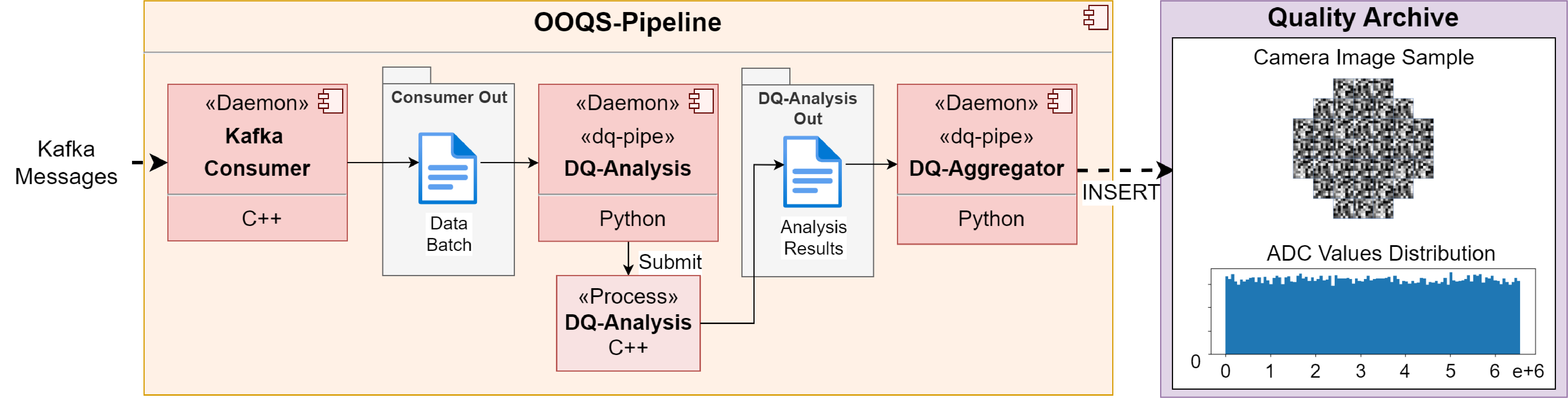}{OOQS_Pipeline_diagram}{The OOQS Pipeline components and the data workflow. Details are in the text.}
It can be scheduled via Slurm, a workload manager for parallel job execution and load distribution in the computing cluster.
The \textit{Kafka Consumer} is connected to the Kafka topic to collect the messages, deserialize camera data using the Avro schema, and collect them in batches. A configurable time window on the data acquisition time determines the size of the data batch. Finally, the application writes the batches into HDF5 files. Data quality analysis relies on two different applications based on the pipeline framework \textit{rta-dq-pipe} \citep{baroncelli2021rtadqlib}: \textit{DQ-Analysis} and \textit{DQ-Aggregator}. 
DQ-Analysis detects the creation of new batch files, and it schedules a process to read the events in the file produced by the consumer, compute histograms, count values, sample images, and write files containing the batch analysis results.
DQ-Aggregator collects the DQ-Analysis results: when a new file is detected, the batch results are read from it, and a stateful class aggregates the results in sliding windows or entire observation collections. The Aggregator inserts the results into the Quality Archive database.  
\section{OOQS Performance}
To evaluate the pipeline performance, we simulated a scientific packet stream of 55 minutes at 1 kHz (3.2M packets, 53 GB). These data were pre-loaded into the Kafka topic. The pipeline prototype supports various analyses, including sampling the camera images, calculating ADC values distributions for each camera module and for the entire camera within a sliding window of 10 seconds, leading to 76 histograms per second, and plotting every second the count of the 37 camera triggers in a sliding window of 10k events.
For comparison, we implemented both Python and C++ Kafka Consumers. The Python consumer processed 600 packets per second, while the C++ version outperformed at 5800 packets/s. Consumer outputs are organized into batch files, each containing data recorded within a 1-second acquisition time slot, with file sizes up to 9.5 MB with 1k events. With nine telescopes, the OOQS server must handle a data write speed of 86 MB/s.
DQ-Analysis yields partial results for use cases on events within 1-second time windows in about 260 ms. DQ-Aggregator updates the analyses sliding windows every second and inserts the result in the quality archive in 30 ms.

\section{Conclusions}
The C++ Kafka consumer efficiently handles data from one telescope at maximum bandwidth, suggesting scalability by deploying multiple Consumers to manage data from multiple telescopes with limited computing resources.
DQ-Analysis and Aggregator efficiently handle batch data and analyses within a fraction of the time window associated with the event batch, making them well-suited for real-time applications. The presented OOQS software will be used in the science verification plan in 2024.
\acknowledgements This work was conducted in the context of the ASTRI Project. We gratefully acknowledge support from the people, agencies, and organizations listed here: http://www.astri.inaf.it/en/library/. This paper went through the internal ASTRI review process.
\bibliography{P202.bib}  
\end{document}